\documentclass[12pt]{article}
\usepackage{amsmath}
\usepackage{amssymb,amscd}
\usepackage{bm}
\def\thefootnote{\fnsymbol{footnote}}
\newlength{\minitwocolumn}
\catcode`\@=11
\long\def\@makefntext#1{
\protect\noindent \hbox to 3.2pt {\hskip-.9pt  
$^{{\eightrm\@thefnmark}}$\hfil}#1\hfill}               %CAN BE USED 

\def\thefootnote{\fnsymbol{footnote}}
\def\@makefnmark{\hbox to 0pt{$^{\@thefnmark}$\hss}}    %ORIGINAL 
        
\def\ps@myheadings{\let\@mkboth\@gobbletwo
\def\@oddhead{\hbox{}
\rightmark\hfil\eightrm\thepage}   
\def\@oddfoot{}\def\@evenhead{\eightrm\thepage\hfil
\leftmark\hbox{}}\def\@evenfoot{}
\def\sectionmark##1{}\def\subsectionmark##1{}}

\font\eightrm=cmr8

\def\MPL{Mod.~Phys.~Lett. }

\def\PL{Phys.~Lett. }
\def\PR{Phys.~Rev. }

\def\PTP{Prog.~Theor.~Phys. }

\newtheorem{thm}{Theorem}[section]

\newtheorem{definition}[thm]{Definition}

\newcommand{\qed}{\nobreak \ifvmode \relax \else
      \ifdim\lastskip<1.5em \hskip-\lastskip
      \hskip1.5em plus0em minus0.5em \fi \nobreak
      \vrule height0.75em width0.5em depth0.25em\fi}

\newcommand\hQ{\mbox{\boldmath $Q$}}

\newcommand\tr{{\rm tr}}

\newcommand{\sbv}[2]{{\{{{#1},{#2}}\}}}
\newcommand{\ssbv}[2]{{\{{{#1},{#2}}\}}}

\newcommand{\bracket}[2]{\langle #1\,,#2\rangle}

\def\bx{\mbox{$x$}}
\def\bxi{\mbox{$\xi$}}
\def\bq{\mbox{$q$}}
\def\bp{\mbox{$p$}}

\def\bbx{\mbox{\boldmath $x$}}
\def\bbxi{\mbox{\boldmath $\xi$}}
\def\bbq{\mbox{\boldmath $q$}}
\def\bbp{\mbox{\boldmath $p$}}
\def\bbe{\mbox{\boldmath $e$}}

\def\bbd{\mbox{\boldmath $d$}}

\def\bomega{\mbox{\boldmath $\omega$}}
\def\brho{\mbox{\boldmath $\rho$}}

\newcommand{\calA}{{\cal A}}

\newcommand{\calC}{{\cal C}}

\newcommand{\calF}{{\cal F}}

\newcommand{\calI}{{\cal I}}

\newcommand{\calL}{{\cal L}}
\newcommand{\calM}{{\cal M}}
\newcommand{\calN}{{\cal N}}
\newcommand{\calO}{{\cal O}}

\newcommand{\calX}{{\cal X}}

\newcommand{\calU}{{\cal U}}

\newcommand{\calW}{{\cal W}}

\newcommand{\Q}{{\kern.24em\vrule width.04em height1.4ex%
                 depth-.05ex\kern-.26em\mathsf Q}}
\newcommand{\C}{{\kern.24em\vrule width.04em height1.4ex%
                 depth-.05ex\kern-.26em\mathsf C}}

\newcommand{\Map}{{\rm Map}}
\newcommand{\ev}{{\rm ev}}
\newcommand{\inte}{{int}}

\begin{document}

%%%%%%%%%%%%%%%%%%%%%%%%%%%%%%%%%%%%%%%%%%%%%%%%%%%%%%%%%%%%%%%%%%
%%%%%%%%%%%%%%%%%%%%%%%% Title %%%%%%%%%%%%%%%%%%%%%%%%%%%%%%%%%%%
%%%%%%%%%%%%%%%%%%%%%%%%%%%%%%%%%%%%%%%%%%%%%%%%%%%%%%%%%%%%%%%%%%

\baselineskip 0.7cm

\begin{titlepage}
%\today
\begin{flushright}
MISC-2011-06
\end{flushright}

\vskip 1.35cm
\begin{center}
{\Large \bf
Donaldson Invariants and Their Generalizations from
AKSZ Topological Field Theories
%in 4 Dimensions
%in Higher Dimensions
%and Topological Invariants
}
\vskip 1.2cm
Noriaki IKEDA%$^1$%
\footnote{E-mail address:\ ikeda@yukawa.kyoto-u.ac.jp}
\vskip 0.4cm
{\it 
%$^1$
Maskawa Institute for Science and Culture,
Kyoto Sangyo University, \\
Kyoto 603-8555, Japan 
%\\
%and \\
%Department of Mathematical Sciences,
%BKC Research Organization of Social Sciences,
%College of Science and Engineering,
%Ritsumeikan University \\
%Kusatsu, Shiga 525-8577, Japan 
}
%{\it
%$^2$
%Sawaya apartments 103,
%Yaraicho 9, Shinjyuku, Tokyo, Japan
%}

\vskip 0.4cm

\today

\vskip 1.5cm

\begin{abstract}
Observable structures of a topological field theory of 
AKSZ type are analyzed.
From a double (or multiple) complex structure 
of observable algebras,
%defined in the AKSZ sigma models and 
new topological invariants are constructed.
Especially, 
Donaldson polynomial invariants
and their generalizations 
are constructed
from a topological field theory of AKSZ type
in $4$ dimensions recently proposed.
\end{abstract}
\end{center}
\end{titlepage}

\renewcommand{\thefootnote}{\alph{footnote}}

\setcounter{page}{2}

%%%%%%%%%%%%%%%%%%%%%%%%%%%%%%%%%%%%%%%%%%%%%%%%%%%%%%%%%%%%%%%%%%
%%%%%%%%%%%%%%%%%%%%%%%% Article %%%%%%%%%%%%%%%%%%%%%%%%%%%%%%%%%
%%%%%%%%%%%%%%%%%%%%%%%%%%%%%%%%%%%%%%%%%%%%%%%%%%%%%%%%%%%%%%%%%%

\rm

%%%%%%%%%%%%%%%%%%%%%%%%%%%%%%%%%%%%%%%%%%%%%%%%%%%%%%%%%%%%%%%%%%%%%
%%%%%%%%%%%%%%%%%%%%%%%%%%%%%%   SEC  1    %%%%%%%%%%%%%%%%%%%%%%%%%%
%%%%%%%%%%%%%%%%%%%%%%%%%%%%%%%%%%%%%%%%%%%%%%%%%%%%%%%%%%%%%%%%%%%%%
\section{Introduction}
\noindent
%and construct their TFTs.
The AKSZ construction 
\cite{Alexandrov:1995kv}\cite{Cattaneo:2001ys}\cite{Roytenberg:2006qz}
is a formulation of a topological field theory (TFT)
based on geometry of a graded symplectic manifold,
so called a QP-manifold.
It is equivalent to the Batalin-Vilkovisky formalism
\cite{Bat}\cite{Schwarz:1992nx} in a gauge theory
if a TFT of Schwarz type is considered.

%\cite{Ikeda:2002wh}\cite{Roytenberg:2006qz}
%\cite{Park:2000au}\cite{Severa:2001}\cite{Ikeda:2001fq}
%\cite{Hansen:2009zd}
A TFT in higher dimensions based in term of a graded manifold has
been formulated in mathematical and physical contexts
\cite{Park:2000au}\cite{Severa:2001}\cite{Ikeda:2001fq}.
The resulting TFTs have structures of homotopy Lie algebroids
and include
known TFTs, such as
BF theories, the Poisson sigma model, the Chern-Simons theory, 
%the Courant sigma model, 
etc.
A review about recent developments 
related to algebroids and topological field theories
are in \cite{Kotov:2010wr}.
%TFTs of AKSZ type in $2$ and $3$ dimensions
%have been analyzed in many papers.
%In this paper, 

An observable structure, which is 
a cohomology class of a complex on a Q-structure,
in a TFT of AKSZ type has been analyzed in many papers
\cite{Schwarz:1999vn}\cite{Lyakhovich:2004kr}\cite{Lyakhovich:2009qq}\cite{Barnich:2009jy}\cite{Qiu:2009rf}\cite{Kallen:2010ff}.
A new aspect of a classical observable
is pointed out in this paper.
Generally a Q-structure has double or multiple
complex structures.
As an interesting case, 
We analyze the AKSZ sigma model 
with a multiple complex structure in this paper.
%,
%which is called a $\hQ_1$ cohomology class,

As an example, a TFT of AKSZ type in $4$ dimensions is considered.
A TFT of AKSZ type in $4$ dimensions has been recently proposed
based on a QP manifold of degree $3$ \cite{Ikeda:2010vz}.
A homotopy algebroid structure in this theory is called 
a Lie algebroid up to homotopy \cite{Ikeda:2010vz} in a splittable case,
or the H-twisted Lie algebroid \cite{Grutzmann} in general case.
This theory includes a BF theory and a topological Yang-Mills theory
in $4$ dimensions as special cases.
Donaldson polynomial invariants \cite{Donaldson}\cite{Witten:1988ze}
are constructed
as a cohomology class of a triple complex of a Q-structure.

The paper is organized as follows. 
In Section 2, a QP-manifold and an AKSZ construction are 
reviewed.
%higher Courant algebroids are reviewed 
%and our notations are fixed.
%defined from the complexs of 
%vector bundles and reformulated by the QP-structures on graded manifolds.
%In section 3, is reviewed.
%$n=\infty$ case is discussed.
In Section 3, a general theory and a multiple complex 
of classical observables of a TFT of AKSZ type is analyzed.
In Section 4, a QP-manifold of degree $3$ is reviewed.
In Section 5, 
Donaldson polynomial invariants are constructed from a 
TFT of AKSZ type
In Section 6, generalizations are considered.
%and 
%by the AKSZ construction of a TFT in $n+1$ dimensions.
%Some examples are listed as special cases.
%In Section 6 is conclusions and discussion.

%%%%%%%%%%%%%%%%%%%%%%%%%%%%%%%%%%%%%%%%%%%%%%%%%%%%%%%%%%%%%%%%%%%%%
%%%%%%%%%%%%%%%%%%%%%%%%%%%%%%%  SEC. 2 %%%%%%%%%%%%%%%%%%%%%%%%%%%%%%
%%%%%%%%%%%%%%%%%%%%%%%%%%%%%%%%%%%%%%%%%%%%%%%%%%%%%%%%%%%%%%%%%%%%
%%%%%%%%%%%%%%%%%%%%%%%%%%%%%%%%%%
\section{QP Manifolds and AKSZ Construction of Topological Field Theories}
%%%%%%%%%%%%%%%%%%%%%%%%%%%%%%%%%%

\begin{definition}
Let $\calM$ be a nonnegatively graded manifold
\footnote{A graded manifold is a ringed space with 
a structure sheaf of nonnegatively graded algebras.},
$\omega$ be a graded symplectic form of degree $n$ on $\calM$ 
and $Q$ be a differential of degree $+1$ with $Q^2=0$.
A triple $(\calM, \omega, Q)$ is called
a \textbf{QP-manifold} of degree $n$ 
if $\omega$ and $Q$ are compatible,
that is, $\calL_Q \omega =0$.
\cite{Schwarz:1992nx}
\end{definition}
The symplectic structure $\omega$
is a called a (classical) \textbf{P-structure} and 
$Q$ is a called a (classical) \textbf{Q-structure}.
The graded Poisson bracket on $C^{\infty}(\calM)$ is 
defined from the graded symplectic structure $\omega$ on $\calM$,
\begin{equation}\label{gradedpoisson}
\sbv{f}{g}=(-1)^{|f|+1}\iota_{X_f}\iota_{X_g}\omega,
\end{equation}
where 
a Hamiltonian vector field $X_f$ is defined by the equation
%\begin{eqnarray}
$
\sbv{f}{g} = X_f g,
$
%\nonumber
%\end{eqnarray}
for $f, g \in C^{\infty}(\calM)$.
%\begin{eqnarray}
%\iota_{X_F} \omega = (-1)^{n+1} d F.
%\nonumber
%\end{eqnarray}
Since 
%$Q$ has the degree $1$ and 
the graded symplectic structure $\omega$ has degree $n$,
the graded Poisson bracket $\sbv{-}{-}$ has degree $-n$.

A Hamiltonian $\Theta\in C^{\infty}(\calM)$ 
of $Q$ with respect to the Poisson bracket $\{-,-\}$
%on $\calM$ 
satisfies
$$
Q=\{\Theta,-\},
$$
and is of degree $n+1$.
The differential condition, $Q^2=0$, implies that
$\Theta$ is a solution of the \textbf{classical master equation},
\begin{equation}\label{CME}
\{\Theta,\Theta\}=0.
\end{equation}

%%%%%%%%%%%%%%%%%%%%%%%%%%%%%%%%%%%%%%%%%%%%%%%%%%%%%%%%%%%%%%%%%%%%%
%%%%%%%%%%%%%%%%%%%%%%%%%%%%%%%  SEC. 4 %%%%%%%%%%%%%%%%%%%%%%%%%%%%%%
%%%%%%%%%%%%%%%%%%%%%%%%%%%%%%%%%%%%%%%%%%%%%%%%%%%%%%%%%%%%%%%%%%%%%
%\section{AKSZ Conctruction of Topological Field Theories
%}
%\noindent
%\medskip\\
%\noindent
If a QP-manifold is given, we can construct a topological field theory 
by the AKSZ construction.
\cite{Alexandrov:1995kv}\cite{Cattaneo:2001ys}\cite{Roytenberg:2006qz}
%%%%%%%%%%%%%%%%%%%%%%%%%%%%%%%%%%%%%%%%%%%%%%%%%%%%%%%%%%%%%%%%%%%%%
%\subsection{QP-Structure on $\Map(\calX, \calM)$}
%\noindent
%In this subsection, w
%We explain the AKSZ construction of 
%a topological field theory in $n+1$ dimensions.

%
%Let $X$ be a manifold in $n+1$ dimensions
%$X$ is called a worldsheet if $n=1$, 
%or a worldvolume if $n >1$.
%and $M$ be a manifold in $d$ dimensions.
%$M$ is a called a target space.
Let $(\calX, D)$ be a differential graded (dg) manifold 
$\calX$
with a $D$-invariant nondegenerate measure $\mu$,
%with $\calX|_0 = X$, 
where
$D$ is a differential on $\calX$.
%($\calM, \Delta, \Theta$)
Let ($\calM, \omega, Q$) be a QP-manifold. 
%$\omega$ is a graded symplectic form of degree $n$ and
$Q= \{\Theta, -\}$ is a differential on $\calM$.
%with $\calM|_0 = M$.
%
%A degree $\deg (-)$ on $\calX$ is called the form degree and 
%a degree $\gh (-)$ on $\calM$ is called the ghost number.
%\footnote{The ghost number $\gh (-)$ is the degree $|-|$
%on $\calM$ in section 2.}.
%
$\Map(\calX, \calM)$ is
%$\bbx:X \longrightarrow M \in 
a space of smooth maps from $\calX$ to $\calM$.
%$|-| = \deg (-) + \gh (-)$ is degree
%on $\Map(\calX, \calM)$ and called the total degree.
%
A QP-structure on $\Map(\calX, \calM)$
is constructed from the above data.

Since 
${\rm Diff}(\calX)\times {\rm Diff}(\calM)$ 
naturally acts on $\Map(\calX, \calM)$,
$D$ and $Q$ induce homological vector fields 
on $\Map(\calX, \calM)$, 
$\hat{D}$ and $\check{Q}$. 
Explicitly, $\hat{D}(z, f) = d f(z) D(z)$ 
and $\check{Q}(z, f) = Q f(x)$,
for $z \in \calX$ and $f \in \calM^{\calX}$.

%
%We consider $\calX = T[1]X$
%In order to construct a QP-structure on $\Map(\calX, \calM)$,

Two maps are introduced.
An {\it evaluation map} 
${\rm ev}: \calX \times \calM^{\calX} \longrightarrow \calM$ 
is defined as
\begin{eqnarray}
{\rm ev}:(z, \Phi) \longmapsto \Phi(z),
\nonumber
\end{eqnarray}
where 
$z \in \calX$ and $\Phi \in \calM^{\calX}$.

A {\it chain map} 
$\mu_*: \Omega^{\bullet}(\calX \times \calM^{\calX}) 
\longrightarrow \Omega^{\bullet}(\calM^{\calX})$ is defined as 
$$\mu_* \omega(y)(v_1, \ldots, v_k)
 = \int_{\calX} \mu(x)
 \omega(x, y) (v_1, \ldots, v_k)$$
where $v$ is a vector field on $\calX$
%$F \in \Omega^{\bullet}(\calX \times \calM)$ 
and 
$\int_{\calX} \mu$ is an integration on $\calX$.
%by the measure $\mu$.
It is an usual integral for even degree parts
and
the Berezin integral for
odd degree parts. 
%such that
%$\int_{\rm fiber} d \theta^{\mu} \ \theta^{\nu} = \delta^{\mu\nu}$, 
%where $\theta^{\mu}$ is a local coordinate on $\calX$.
%\noindent
%{\bf Definition}

A P-structure on $\Map(\calX, \calM)$ is defined as follows:
\begin{definition} 
For a graded symplectic form $\omega$ on 
%For an odd Laplace operator $\Delta$ on 
$\calM$, 
a graded symplectic form $\bomega$ on $\Map(\calX, \calM)$ is
defined as
$\bomega := \mu_* \ev^* \omega$,
%where $F \in C^{\infty}(\calM)$.
%a linear operator ${\bDelta}$ on $\Map(\calX, \calM)$ as
%${\bDelta} \mu_* \ev^* F = \mu_* \ev^* \Delta F$,
where $F, G \in C^{\infty}(\calM)$.
\end{definition}
%
%We can confirm that 
$\bomega$ is nondegenerate and closed
because $\mu_* \ev^*$ preserves nondegeneracy and closedness.
A graded symplectic form $\bomega$ 
%is a graded symplectic form on $\Map(\calX, \calM)$ and 
induces a (graded) Poisson bracket 
$\ssbv{-}{-}$
on $\Map(\calX, \calM)$.
$|\bomega| = -1$ because $|\omega|=n$ and
$|\mu_* \ev^*|=-n-1$.
Therefore a Poisson bracket $\ssbv{-}{-}$ 
on $\Map(\calX, \calM)$ has degree $1$ and odd.
%
% on $\Map(\calX, \calM)$, 

An odd Laplace operator ${\Delta}$
on $\Map(\calX, \calM)$ can be constructed
from an odd Poisson bracket 
if $\Map(\calX, \calM)$ has a measure $\brho$.
%by a similar method as in section 4.1.
That is, it is defined as
\begin{eqnarray}
{\Delta} F = \frac{(-1)^{|F|}}{2} {\rm div}_{\brho} X_F,
\nonumber
\end{eqnarray}
where $F \in C^{\infty}(\Map(\calX, \calM))$.
Here the divergence ${\rm div}$ is defined as
$\int_{\calM} \brho \ X(F) = \int_{\calM} \brho \ {\rm div}_{\rho} X F$ 
for $F \in C^{\infty}(\Map(\calX, \calM))$.
%
%An odd Laplace operator is necessary for a BV quantization of a TFT.
%
Conversely, If an odd Laplace operator is defined on $\Map(\calX, \calM)$,
an odd Poisson bracket is derived from an odd Laplace operator 
by a derived bracket:
\begin{eqnarray}
\{F,G\}:=-(-1)^{|F|}[[\hat{\Delta},F],G](1).
\nonumber
%\label{oddlaplacian}
\end{eqnarray}

%$\Delta^2 = 0$
%odd Laplace operator 
%($\bb$)
%A
%
%
%
%$F, G \in Hom(\sum_k C^{\infty}(M)^{\otimes k}, C^{\infty}(M))$
%\begin{eqnarray}
%$
%\sbv{F}{G} 
%= \Omega^{-1} \sbv{F}{G} 
%&=& 
%\equiv (-1)^{(n+1) |F|} \Delta(FG) - (-1)^{(n+1) |F|} \Delta(F){G} 
%- (-1)^{n|F|} {F}\Delta(G)
%$
%\nonumber 
%\label{Poisson2ike}
%\end{eqnarray}
%}
%$\sbv{F}{G}$
%degree $-n+1$ Loday bracket
%

A Q-structure $S$ on $\Map(\calX,  \calM)$
$S$ is called a {\it BV action} and
consists of two parts $S = S_0 + S_{\inte}$.
$S$ is constructed as follows.
%
%$S_0$ is constructed as follows: 
%Let 
%$\sbv{\cdot}{\cdot}$ be
%a nondegenerate graded Poisson bracket 
%induced from a P-structure on $\calM$. 
%This defines t
%Let $\omega$ be the odd symplectic form derived from 
%a P-structure on $\calM$.
%
%(= D \bphi \wedge D \bb_1 -D \ba_1 \wedge D \ba_1)$ 
%($D=\theta^{\mu} \partial_{\mu}$) 
We take a fundamental form $\vartheta$ such that 
$\bomega= - d \vartheta$ and
%$\Xi = {\rm ev}^* \vartheta$
define $S_0 := \iota_{\hat{D}} \mu_* {\rm ev}^* \vartheta$.
%
%$S_1$ is constructed as follows: 
%
We take a Q-structure $\Theta$ on $\calM$ and 
%A homological vector field 
define $S_{\inte} := \mu_* \ev^* \Theta$.
$|S_0|=0$ and $S_{\inte}$ have degree $0$
because $\mu_* {\rm ev}^*$ has degree $-n-1$.
% because of $\mu_* {\rm ev}^*$.
%$|\Theta|=n \Longleftrightarrow |S|=0$.

We can prove that
$S$ is a Q-structure on $\Map(\calX,  \calM)$, 
since 
\begin{eqnarray}
\sbv{\Theta}{\Theta} =0
\Longleftrightarrow \ssbv{S}{S} =0.
\label{classicalmaster}
\end{eqnarray}
from the definition of $S_0$ and $S_1$.
The right hand side in this equation is called
the classical master equation
on $\Map(\calX, \calM)$.
A homological vector field $\hQ = \sbv{S}{-}$ has
$|\hQ|=1$.
%since degree of $\sbv{-}{-}$ and $S$.
$\hQ$ is a coboundary operator which defines a cohomology,
called a BRST cohomology.
%because $\hQ^2=0$ from the classical master equation.
%

%Note that $S_0$ is defined locally, but 
%(\ref{classicalmaster}) can be confirmed globally.

%The infinitesimal form of the right hand side 
%in (\ref{classicalmaster}) is 
%$\sbv{S}{S} - 2 i \hbar \hat{\bDelta} S =0$.
%This equation is called a {\it quantum master equation}.
%We summarize 
%Then t
The following theorem has been proved
\cite{Alexandrov:1995kv}:
\begin{thm}
If $\calX$ is a dg manifold with a compatible measure and 
$\calM$ is a QP-manifold,
%with $\calM|_0=M$,
the graded manifold $\Map(\calX, \calM)$
has a QP-structure.
%induced from the QP-structure on $\calM$.
\end{thm}

%In $\hbar \longrightarrow 0$ (classical) limit,
%(\ref{classicalmaster}) reduces to

%We define a quadraple ($\calX, \calM, \hat{\bDelta}, S$)
%as {\it a topological field theory},
%induced from 
%constructed in the theorem from the above construction
%induced from a QP-structure on $\calM$.
%
%\noindent
%{\bf Definition}:
This geometric structure on $\Map(\calX,  \calM)$
is called a topological field theory.
A QP-structure is the consistency condition of a topological field theory.

For a graded manifold $\calN$,  
let $\calN|_0$ be the degree zero part.
\begin{definition}
A \textbf{topological field theory} in $n+1$ dimensions
is a triple ($\calX, \calM, S$), 
where $\calX$ is a dg manifold with
$\dim \calX|_0 = n+1$, 
$\calM$ is 
a QP-manifold of degree $n$
%$\bomega$ is a graded symplectic form of degree $n$
and $S$ is a BV action of degree $0$ such that 
$\ssbv{S}{S} =0$.
%$\hat{\bDelta} (e^{\frac{i}{\hbar} S}) =0$.
\end{definition}

%on a $\bZ_n$ graded manifold $\Map(\calX, \calM)$.
%Let $\calX, \calM$ be a graded manifold and
%$\dim \calX|_0 =n$.
%, where $\calX|_0$ is a degree $0$ part of $\calX$.

%We define a P-structure on $\Map(\calX \calM)$.
%In order to interpret this theory
%as a `physical' topological field theory,
%we take $\calX= 
%$ a special graded manifold, $
%T[1]X$.
%and $\calM$ is a $\bZ_n$ graded manifold.
%, so far.
If $\calX = T[1]X$, 
%we can prove that 
a QP-structure on $\Map(\calX, \calM)$ is 
equivalent to the BV formalism of a topological field theory
\cite{Cattaneo:2001ys}\cite{Ikeda:2001fq}.
%\footnote{The AKSZ formalism is generalized to realize
%a BFV formalism if $\calX$ has degree $n$.
%\cite{Cattaneo:2010re}
%}
%We can confirm that this 
The theory is gauge invariant and unitary
by physical discussion.
Thus it provides a consistent quantum field theory. 
%That is, this 
%quantum field theory
%is gauge invariant, unitary
%and has no gauge anomaly.
%If $n \leq 3$, it is renormalizable.

%We set $\calX = T[1]X$ from now.

%Physically we can prove that
%a QP-manifold 
%(or a $\Delta$-cohomology) 
%with the degree $n$
%on a $\bZ_n$ graded manifold $\Map(\calX, \calM)$
%is equivalent the AKSZ-BV formalism of a topological field theory

%The quantization procedure is following.
%(\ref{divergence})
%(\ref{divergence2})

$
\calA = 
\sbv{S_{\inte}}{-}
%- \frac{S_1 \rd}{\partial \bbxi_j}\frac{\ld}{\partial \bbx^j} 
$
is a generalized connection on $\calX \otimes \bbx^*(\calM)$
and 
$\hQ \bbe_{a(i)} = (\bbd + \calA) \bbe_{a(i)}$ is a covariant derivative,
where $\bbe_{a(i)}$ is a basis of degree $i$ elements.
In fact, this coincides with a connection form,
if a graded manifold constructed from 
a principal bundle with a structure Lie group $G$
and 
a nonabelian BF Theory in $n+1$ dimensions
are considered.
Since $\hQ^2 F(\bbe) = (\bbd + \calA) (\bbd + \calA) F(\bbe)
= \calF F(\bbe)$,
$\hQ^2=0$ is equivalent to the flatness condition $\calF=0$,
where $\calF = \bbd \calA + \calA \calA =0$ 
is a curvature.
%Since $\sbv{S_0}{S_0} = 0$, $S_0$ is considered a 'differential',
%and $S_1$ a 'connection'.
The classical master equation $\ssbv{S}{S} =0$ is 
%a sort of 
%a flatness condition, that is, 
considered as the Maurer-Cartan equation.

%%%%%%%%%%%%%%%%%%%%%%%%%%%%%%%%%%%%%%%%%%%%%%%%%%%%%%%%%%%%%%%%%%
%%%%%%%%%%%%%%%%%%%%%%%%%%%%%%%%%%%%%%%%%%%%%%%%%%%%%%%%%%%%%%%%%%
%%%%%%%%%%%%%%%%%%%%%%%%%%%%%%%%%%%%%%%%%%%%%%%%%%%%%%%%%%%%%%%%%%
\section{Double and Triple Complexes and Cohomology}
\noindent
In the AKSZ construction, 
$\hQ$ is decomposed to two coboundary operators
$\hQ = \hQ_0 + \hQ_{\inte}$
%$\hhQ = \hQ^{(0)} + \hQ^{(1)} + \hQ^{(2)}$
such that 
\begin{eqnarray}
&& \hQ_0^2 = \hQ_{\inte}^2 = 0, 
\nonumber \\
&& \sbv{\hQ_0}{\hQ_{\inte}} = 0, \qquad\qquad  \mbox{if} \qquad k \neq l,
\nonumber 
\end{eqnarray}
where $\hQ_0 = \sbv{\hQ_0}{-}$ and $\hQ_{\inte} = \sbv{\hQ_{\inte}}{-}$
$\hQ_0$ and $\hQ_{\inte}$ defines a double complex 
$\calC^{p_0,p_1}(\hQ_0, \hQ_{\inte})$.
%We take $\hQ =\bbd$, the differential on $\calX$.

%%%%%%%%%%%%%%%%%%%%%%%%%%%%%%%%%%%%%%%%%%%%%%%%%%%%%%%%%%%%%%%%%%%%%
%\subsection{$\hQ_1$-Cohomology}
%\noindent
%In the AKSZ formulation of a TFT, 
%if we take
%$\hQ^{(0)} = \hQ_0 = \bbd$ and
%$\hQ^{(1)} = \hQ_1$,
%these define a double complex 
%$\calC^{p_0,p_1}(\bbd, \hQ)$
%and 
%a coboundary operator 
%$\hhQ = \hQ = \hQ_0 + \hQ_1$
%since 
%\begin{eqnarray}
%&& \hQ^2 = \hQ_0^2 = 0, 
%\nonumber \\
%&& \sbv{\hQ}{\hQ_0} = 0,
%\nonumber 
%\end{eqnarray}
We are interested in a $\hQ$-cohomology class.
%which is defined from the Q-structure on $\calM$.
Let us take a $\hQ_{\inte}$-cocycle 
$\calW$ such that $\hQ_{\inte} \calW 
= (\hQ - \hQ_0) \calW = 0$.
$\calW$ is constructed from $Q$-cocycle $W$ on $\calM$ as
$\bbx^* W$.
Then 
since $\hQ \calW = \hQ_0 \calW = \bbd \calW$,
\begin{thm}\label{q1cohomology}
\begin{eqnarray}
\mu_{k*} \calW = \int_{\gamma_k} \mu_k \ \calW,
\nonumber
\end{eqnarray}
is a $\hQ$-cocycle, where $\gamma_k$ is a $k$-cycle on $\calX$
and $\mu_k$ is an induced measure of $\mu$ on $\gamma_k$.
\end{thm}
$\calW$ is expanded by $\theta$
as
\begin{eqnarray}
\calW = \sum_{k=0}^{n+1} \calW_k (\sigma, \theta)
= \sum_{k=0}^{n+1} 
\theta^{\mu(1)}\cdots \theta^{\mu(k)}
\calW_{k, \mu(1) \cdots \mu(k)}(\sigma).
\nonumber
\end{eqnarray}
By substituting this expansion to the equation 
$\hQ_1 \calW =0$,
%(\ref{QWN}), 
the following descent equations are obtained,
\begin{eqnarray}
\hQ \calW_0 =0, 
\nonumber \\
- \bbd \calW_1 + \hQ \calW_0 =0, 
\nonumber \\
\vdots
\nonumber \\
- \bbd \calW_{n} + \hQ \calW_{n-1} =0, 
\nonumber \\
- \bbd \calW_{n} =0, 
\nonumber
\end{eqnarray}
where $\hQ_0=\bbd$ is used.
The following theorem is obtained:
\begin{thm}\label{kcocycle}
The integration of $\calW_k$
on a $k$-cycle $\gamma_k$ of $\calX$,
\begin{eqnarray}
\calO_{k} = \int_{\gamma_k} \mu_k \ \calW
= \int_{\gamma_k} \mu_k \ \calW_k 
\nonumber
\end{eqnarray}
is a $\hQ$-cohomology class,
\end{thm}
by applying the theorem \ref{q1cohomology} to $\calW_k$.

Especially, if we consider 
a $\hQ$-exact cocycle:
%a $\hQ_{\inte}$-exact cocycle:
\begin{eqnarray}
\calW = \hQ \calU, 
%\qquad 
%\calW = \hQ_{\inte} \calU,
%\label{calfdef}
\end{eqnarray}
%$\hQ \calW = 0$ is trivially satisfied and 
$\mu_{*k} \calW$ is a $\hQ$-cocycle.
Some important geometric objects have these forms.
Let $\bbx$ be a map $\bbx: \calX \longrightarrow M$, 
where $M = \calM_0$ be a manifold of a degree $0$ 
part of $\calM$.
If we consider $\calU_0 = F(\bbx)$, a function of $\bbx$, 
$\hQ$ acts as a bundle map (the anchor map)
$\rho:E \longrightarrow TM$ of an algebroid.
\begin{eqnarray}
\mu_{*1} \calW
%+ \frac{S_1 \rd}{\partial \bbe_{b(j)}}
%\omega_{b(j)a(k)}
%\frac{\ld}{\partial e_{a(k)}} e_{a(i)}
= \int_{\gamma_1} \mu_1 \ \hQ F(\bbx),
%= \mu_{*1} \ev^* X_f
%= 
%\int_{\gamma_1} \mu \ 
%\frac{\partial F}{\partial \bbx^i} f_{1}{}^{i}{}_{a}(\bbx) \bbq^{a},
\nonumber
\end{eqnarray}
is a Hamiltonian vector field with respect to $\hQ$.
%where $F = \ev^* f$.
%
Let $\bbq^a$ be a section of $T^*[1]X \otimes \bbx^*(\calM_1)$,
where $\calM_1$ is a degree one part of $\calM$.
If we consider $\calU = \bbq^a$,
$\hQ \bbq^a$ is a generalized curvature.
The Bianchi identity is $\hQ^2 \bbq^a =0$.
\begin{eqnarray}
\mu_{*2} \calW
%+ \frac{S_1 \rd}{\partial \bbe_{b(j)}}
%\omega_{b(j)a(k)}
%\frac{\ld}{\partial e_{a(k)}} e_{a(i)}
= \int_{\gamma_2} \mu_2 \ \hQ (U_a(\bbx) \bbq^a),
%= 
%\int_{\gamma_2} \mu \ 
%\frac{\partial F}{\partial \bbx^i} f_{1}{}^{i}{}_{a}(\bbx) \bbq^{a},
\nonumber
\end{eqnarray}
is a $\hQ$ cohomology class.
This is a generalization of the $1$st Pontryagin class.
\medskip\\
\noindent
%For $\sum_{k=0}^n \hQ^{(k)}$,
We assume that a Q-structure
$\hQ$ is decomposed to three coboundary operators
$\hQ = \hQ^{(0)} + \hQ^{(1)} + \hQ^{(2)}$
such that 
\begin{eqnarray}
&& (\hQ^{(k)})^2 = 0, 
\nonumber \\
&& \sbv{\hQ^{(k)}}{\hQ^{(l)}} = 0, \qquad\qquad  \mbox{if} \qquad k \neq l,
\nonumber 
\end{eqnarray}
where $\hQ^{(0)} = \hQ_0$ and $\hQ_{\inte} = \hQ^{(1)} + \hQ^{(2)}$.
This defines a triple complex $\calC^{p_0,p_1,p_2}(\hQ^{(k)})$.

In this case, 
a $\hQ^{(2)}$-exact cocycle:
\begin{eqnarray}
\calW = \hQ^{(2)} \calU,
\label{calfdef}
\end{eqnarray}
is nontrivial $\hQ$-cocycle
if $\calW$ satisfies $\hQ^{(1)} \calW = 0$ 
and $\calW$ is not $\hQ^{(1)}$-exact,
where $\calU \in C^{\infty}(\Map(\calX, \calM))$.
%$\mu_{*k} \calW$ is a $\hQ$-cocycle
%$\calF$ trivially satisfies 
%since 
%the theorem \ref{q1cohomology} and
%$\hQ^{(2)} \calW = 0$ is trivially satisfied.
%This cocylce is not $\hQ_1$-exact.
%Let us take
%$\calU = \bbe_{a(i)}$.
%Then 
%$$
%\calW 
%%= \hQ \calG 
%= \hQ \bbe_{a(i)} 
%= \bbd \bbe_{a(i)} 
%+ \sbv{S_1}{\bbe_{a(i)}}
%= \bbd \bbe_{a(i)} 
%- \frac{S_1 \rd}{\partial \bbe_{b(j)}} \bomega_{b(j)a(i)}
%$$
%Especially, 
%$\calG = \bbx^{a(0)} = \bbx^{i}$,
%where we denote $a(0)=i$.
%Then
%$$
%\calF = \hQ \bbx^{i} = \bbd \bbx^i 
%- \frac{S_1 \rd}{\partial \bbxi_j}\frac{\ld}{\partial \bbx^j} \bbx^i
%= \bbd \bbx^i 
%- \frac{S_1 \rd}{\partial \bbxi_i}.
%$$

%%%%%%%%%%%%%%%%%%%%%%%%%%%%%%%%%%%%%%%%%%%%%%%%%%%%%%%%%%%%%%%%%%%%%
%%%%%%%%%%%%%%%%%%%%%%%%%%%%%%%  SEC. 5 %%%%%%%%%%%%%%%%%%%%%%%%%%%%%%
%%%%%%%%%%%%%%%%%%%%%%%%%%%%%%%%%%%%%%%%%%%%%%%%%%%%%%%%%%%%%%%%%%%%%
\section{QP Structures of Degree $3$}
\noindent
In this section, we remember a QP-structure of degree $3$
and a Lie algebroid up to homotopy introduced
in the paper \cite{Ikeda:2010vz}.

\subsection{P-structures}
\noindent
Let $E\to M$ be a vector bundle over an ordinary smooth manifold $M$.
The shifted bundle $E[1]\to M$ is a graded manifold
whose fiber space has degree $+1$.
We consider the shifted cotangent bundle $\calM:=T^{*}[3]E[1]$.
It is a P-manifold of the degree $3$ over $M$,
$$
T^{*}[3]E[1]\to\calM_{2}\to E[1]\to M.
$$
%where %$\calM_{2}$ is a certain graded manifold
%\footnote{
In fact $\calM_{2}$ is $E[1]\oplus E^*[2]$.
%which is derived from the result
%in the previous sentence of Remark 3.2.
%}.
The structure sheaf $C^{\infty}(\calM)$ of $\calM$
is decomposed into the homogeneous subspaces,
$$
C^{\infty}(\calM)=\sum_{i \ge 0}C^{i}(\calM),
$$
where $C^{i}(\calM)$ is the space of functions of degree $i$.
In particular, $C^{0}(\calM)=C^{\infty}(M)$: the algebra of
smooth functions on the base manifold
and $C^{1}(\calM)=\Gamma E^{*}$: the space
of sections of the dual bundle of $E$.
We have
$
C^{2}(\calM)=\Gamma E\oplus\Gamma\wedge^{2}E^{*}.
$
\\
\indent
Let us denote by $(x,q,p,\xi)$
a canonical (Darboux) coordinate on $\calM$, where
$x$ is a smooth coordinate on $M$,
$q$ is a fiber coordinate on $E[1]\to M$,
$(\xi,p)$ is the momentum coordinate
on $T^{*}[3]E[1]$ for $(x,q)$.
The degrees of the variables $(x,q,p,\xi)$ are respectively $(0,1,2,3)$.\\
\indent
%Since $C^{2,0}(\calM)=\Gamma E$ and $C^{0,2}(\calM)=\Gamma\wedge^{2}E^{*}$,
For the canonical coordinate on $\calM$,
the symplectic structure has the following form:
\begin{eqnarray}\label{bomega}
\omega = \delta \bx^i \delta \bxi_i + \delta \bq^a \delta \bp_a,
\end{eqnarray}

%%%%%%%%%%%%%%%%%%%%%%%
\subsection{Q-structures}
%%%%%%%%%%%%%%%%%%%%%%%%%
\begin{definition}
%(Bidegree, 
(see also Remark 3.3.3 in \cite{Roy01})
Consider a monomial $\xi^{i}p^{j}q^{k}$ on a local chart
$(U;x,q,p,\xi)$ of $\calM$, of which the total degree is $3i+2j+k$.
The \textbf{bidegree} of the monomial is, by definition,
$(2(i+j),i+k)$.
\end{definition}
The bidegree is globally well-defined.
%(See also Remark \ref{shiftremark} below.)\\
We consider a Q-structure, 
$Q= \sbv{\Theta}{-}$, on the P-manifold.
It is required that $\Theta$ has degree $4$.
That is, $\Theta\in C^{4}(\calM)$.
Because $C^{4}(\calM)=C^{4,0}(\calM)\oplus 
C^{2,2}(\calM)\oplus C^{0,4}(\calM)$,
the Q-structure is uniquely decomposed into
$$
\Theta=\theta_{2}+\theta_{13}+\theta_{4},
$$
%where the bidegrees of the substructures are
%$(4,0)$, $(2,2)$ and $(0,4)$, respectively.
In the canonical coordinate, $\Theta$ is the following polynomial:
\begin{eqnarray}\label{deftheta}
\Theta = f{}_1{}^i{}_{a} (\bx) \bxi_i \bq^a 
+\frac{1}{2} f_2{}^{ab}(\bx) \bp_a \bp_b
+\frac{1}{2} f_3{}^a{}_{bc}(\bx) \bp_a \bq^b \bq^c
+\frac{1}{4!} f_4{}_{abcd}(\bx) \bq^a \bq^b \bq^c \bq^d,
\end{eqnarray}
and the substructures are 
\begin{eqnarray*}
\theta_{2}&=&\frac{1}{2} f_2{}^{ab}(\bx) \bp_a \bp_b,\\
\theta_{13}&=&
f{}_1{}^i{}_{a} (\bx) \bxi_i \bq^a+\frac{1}{2} f_3{}^a{}_{bc}(\bx) \bp_a \bq^b \bq^c,\\
\theta_{4}&=&\frac{1}{4!} f_4{}_{abcd}(\bx) \bq^a \bq^b \bq^c \bq^d,
\end{eqnarray*}
%respectively, 
where $f_{1}$-$f_{4}$ are structure functions on $M$.
%By  counting the bidegree, 
One can easily prove that
the classical master equation $\sbv{\Theta}{\Theta} = 0$
is equivalent to the following three identities:
\begin{eqnarray}
\label{tc1}
\{\theta_{13},\theta_{2}\}&=&0,\\
\label{tc2}
\frac{1}{2}\{\theta_{13},\theta_{13}\}
+\{\theta_{2},\theta_{4}\}&=&0,\\
\label{tc3}
\{\theta_{13},\theta_{4}\}&=&0.
\end{eqnarray}
%The conditions (\ref{tc1}), (\ref{tc2}) and (\ref{tc3})
%are equivalent to
%\begin{eqnarray}
%\label{fc1}
%&&f{}_1{}^i{}_{b} f_2{}^{ba} = 0,\\
%\label{fc2}
%&&
%f{}_1{}^k{}_{c} \frac{\partial f_2{}^{ab}}{\partial x^k} 
%+ f_2{}^{da} f_3{}^b{}_{cd} + f_2{}^{db} f_3{}^a{}_{cd} = 0,\\
%\label{fc3}
%&&
%f{}_1{}^k{}_{b} \frac{\partial f{}_1{}^i{}_{a}}{\partial x^k} 
%- f{}_1{}^k{}_{a} \frac{\partial f{}_1{}^i{}_{b}}{\partial x^k} 
%+ f{}_1{}^i{}_{c} f_3{}^c{}_{ab} = 0,\\
%\label{fc4}
%&& 
%f{}_1{}^k{}_{[d} \frac{\partial f_3{}^a{}_{bc]}}{\partial x^k} 
%+ f_2{}^{ae} f_4{}_{bcde}
%- f_3{}^a{}_{e[b} f_3{}^e{}_{cd]} = 0,\\
%\label{fc5}
%&& 
%f{}_1{}^k{}_{[a} \frac{\partial f_4{}_{bcde]}}{\partial x^k} 
%+ f_3{}^f{}_{[ab} f_4{}_{cde]f} =0,
%\end{eqnarray}
%where $[b \ c \ d \ \cdots]$ is a skewsymmetrization
%with respect to indices $b, c, d, \cdots$, etc.
%%%%%%%%%%%%%%%%%%%%%%%%%%%%%%%%%
\subsection{Lie algebroid up to homotopy}
%%%%%%%%%%%%%%%%%%%%%%%%%%%%%%%%%
%In this section we study an algebraic structure
%associated with the QP-structure in 5.1 and 5.2.
\begin{definition}
Let $Q=\theta_{2}+\theta_{13}+\theta_{4}$ be a $Q$-structure
on $T^{*}[3]E[1]$, where $(\theta_{2},\theta_{13},\theta_{4})$
is the unique decomposition of $\Theta$.
We call the quadruple $(E; \theta_{2},\theta_{13},\theta_{4})$
a \textbf{Lie algebroid up to homotopy},
%or shortly, 
in shorthand, Lie algebroid u.t.h.
\end{definition}
%We should study the algebraic properties of the Lie algebroid up to homotopy.
Let us define a bracket product by
\begin{equation}\label{braee}
[e_{1},e_{2}]:=\{\{\theta_{13},e_{1}\},e_{2}\},
\end{equation}
where $e_{1},e_{2}\in\Gamma E$.
%By the bidegree counting, 
%one can easily check that
%$\Gamma E$ is closed under this bracket.
The bracket is not necessarily a Lie bracket,
but it is still skewsymmetric:
%\begin{eqnarray*}
%[e_{1},e_{2}]&=&\{\{\theta_{13},e_{1}\},e_{2}\},\\
%&=&\{\theta_{13},\{e_{1},e_{2}\}\}+\{e_{1},\{\theta_{13},e_{2}\}\},\\
%&=&-\{\{\theta_{13},e_{2}\},e_{1}\}=-[e_{2},e_{1}],
%\end{eqnarray*}
%where $\{e_{1},e_{2}\}=0$ is used.
A bundle map $\rho:E\to TM$ which is called an anchor map
is defined by the following identity:
% below,
$$
\rho(e)(f):=\{\{\theta_{13},e\},f\},
$$
where $f\in C^{\infty}(M)$.
The bracket and the anchor map 
satisfy the \textbf{algebroid} conditions (A0) and (A1) below:
\begin{description}
\item[(A0)]
$\rho[e_{1},e_{2}]=[\rho(e_{1}),\rho(e_{2})]$,
\item[(A1)]
$[e_{1},fe_{2}]=f[e_{1},e_{2}]+\rho(e_{1})(f)e_{2}$,
\end{description}
where the bracket $[\rho(e_{1}),\rho(e_{2})]$
is the usual Lie bracket on $\Gamma TM$.
The structures $\theta_{13}$, $\theta_{2}$ and $\theta_{4}$ define
the three operations:
\begin{itemize}
\item $\delta(-):=\{\theta_{13},-\}$;
a de Rham type derivation on $\Gamma\wedge^{\cdot}E^{*}$,
\item $(\alpha_{1},\alpha_{2}):=\{\{\theta_{2},\alpha_{1}\},\alpha_{2}\}$;
a symmetric pairing on $E^{*}$, where $\alpha_{1},\alpha_{2}\in\Gamma E^{*}$,
\item $\Omega(e_{1},e_{2},e_{3},e_{4}):=
\{\{\{\{\{\theta_{4},e_{1}\},e_{2}\},e_{3}\},e_{4}\}$;
a 4-form on $E$.
\end{itemize}
%Remark that $\delta\delta\neq 0$ in general.
%Because the degree of the pairing is $-2$,
%it is $C^{\infty}(M)$-valued.
The pairing induces a symmetric
bundle map $\partial:E^{*}\to E$
which is defined by the equation,
$(\alpha_{1},\alpha_{2})=\bracket{\partial \alpha_{1}}{\alpha_{2}}$,
where $\bracket{-}{-}$ is the canonical pairing
of the duality of $E$ and $E^{*}$.
Since $\bracket{\alpha}{e}=\{\alpha,e\}$, we have
$$
\partial\alpha=-\{\theta_{2},\alpha\}.
$$
%From Eq.~(\ref{tc2}), w
We get an explicit formula of the Jacobi anomaly
from $\sbv{\Theta}{\Theta}=0$:
\begin{description}
\item[(A2)]
$[[e_{1},e_{2}],e_{3}]+({\rm cyclic \ permutations})
= \partial\Omega(e_{1},e_{2},e_{3})$.
\end{description}
In a similar way, we obtain the following identities:
\begin{description}
\item[(A3)] $\rho\partial=0$,
\item[(A4)] $\rho(e)(\alpha_{1},\alpha_{2})=(\mathcal{L}_{e}\alpha_{1},\alpha_{2})
+(\alpha_{1},\mathcal{L}_{e}\alpha_{2})$,
\item[(A5)] $\delta\Omega=0$,
\end{description}
where $\mathcal{L}_{e}(-):=\{\{\theta_{13},e\},-\}$
is the Lie type derivation which acts on $E^{*}$.
%Axioms (A3) and (A4) are induced from Eq.~(\ref{tc1})
%and (A5) is from Eq.~(\ref{tc3}).
\\
\indent
%The fundamental relations (\ref{fc1})--(\ref{fc5})
%correspond to Axioms (A1)--(A5)
Thus, 
%the notion of 
the Lie algebroid up to homotopy
is characterized by the algebraic properties (A1)--(A5)
\footnote{
Actually, the axiom (A0) depends on (A1) and (A2).
}.
One concludes that
%\medskip\\
%\noindent
%{\em
\begin{thm}
The classical algebra associated with the QP-manifold $(T^{*}[3]E[1],\Theta)$
is the space of sections of the vector bundle $E$ with the operations
$([\cdot,\cdot],\rho,\partial,\Omega)$
satisfying (A1)--(A5).
\end{thm}
%}
%\medskip\\
%\indent
%In the next section, we will study some special examples
%of Lie algebroid u.t.h.s.

%%%%%%%%%%%%%%%%%%%%%%%%%%%%%%%%%%%%%%%%%%%%%%%%%%%%%%%%%%%%%%%%%%%%%
%%%%%%%%%%%%%%%%%%%%%%%%%%%% SEC. 6 %%%%%%%%%%%%%%%%%%%%%%%%%%%%%%%%%
%%%%%%%%%%%%%%%%%%%%%%%%%%%%%%%%%%%%%%%%%%%%%%%%%%%%%%%%%%%%%%%%%%%%%
\section{Donaldson Polynomials from AKSZ TFT in $4$ Dimensions}
\noindent
A general AKSZ construction in section 2
is applied to a QP manifold of degree $3$.
%the local coordinate expression in section 3.1 and 
%and a topological field theory in $4$ dimensions.
%is obtained as a special case and 
%A new nontrivial topological field theory is constructed.
Let us take a manifold $X$ in $4$ dimensions and
a manifold $M$ in $d$ dimensions.
For a graded vector bundle $E[1]$ on $M$,
take $\calX = T[1]X$
and $\calM = T^*[3]E[1]$.

$(\sigma^{\mu}, \theta^{\mu})$ is a local coordinate
on $T[1]X$. $\sigma^{\mu}$ is a local coordinate on 
the base manifold $X$ of degree zero and 
$\theta^{\mu}$ is one on the fiber of $T[1]X$ of degree $1$, 
respectively.
$\bbx^i$ is a smooth map $\bbx^i: X \longrightarrow M$ and 
$\bbxi_i$ is a section of $T^*[1]X \otimes \bbx^*(T^*[3] M)$, 
$\bbq^a$ is a section of $T^*[1]X \otimes \bbx^*(E[1])$ and
$\bbp_a$ is a section of $T^*[1]X \otimes \bbx^*(T^*[3]E[1])$.
The exterior derivative $d$ is taken 
as a differential $D$ on $X$.
These are called {\it superfields}.
From $d$, a differential 
$\bbd = \theta^{\mu} \frac{\partial}{\partial \sigma^{\mu}}$
on $\calX$ is induced.

%In a local coordinate, 
Then a BV action $S$ has the following expression from 
(\ref{bomega}) and (\ref{deftheta}):
\begin{eqnarray}
S&=&S_0 + S_{\inte},
\label{4DBVaction} \\
S_0&=&\int_{\calX} d^4\sigma d^4\theta (\bbxi_i \bbd \bbx^i
- \bbp_a \bbd \bbq^a),
\nonumber \\
S_{\inte}&=&
\int_{\calX} d^4\sigma d^4\theta 
\left(f{}_1{}^i{}_{a} (\bbx) \bbxi_i \bbq^a 
+ \frac{1}{2} f_2{}^{ab}(\bbx) \bbp_a \bbp_b
+ \frac{1}{2} f_3{}^a{}_{bc}(\bbx) \bbp_a \bbq^b \bbq^c
+ \frac{1}{4!} f_4{}_{abcd}(\bbx) \bbq^a \bbq^b \bbq^c \bbq^d
\right).
\nonumber
\end{eqnarray}
where $\mu = d^4\sigma d^4\theta$.
We introduce a tridegree.
\begin{definition}
%(Bidegree, 
Consider a monomial $\bbxi^{i} \bbp^{j} \bbq^{k}
\bbx^{l}\sigma^{m}\theta^{n}$ on 
a local chart of $\Map(\calX, \calM)$,
of which the total degree is $3i+2j+k+n$.
The \textbf{tridegree} of the monomial is, by definition,
$(2(i+j),i+k, n)$.
\end{definition}
First two degrees in tridegree are ones 
induced from bidegree on $\calM$.
$S_{\inte}$ is decomposed to three terms as
\begin{eqnarray*}
S_{13}&=&
\int_{\calX} d^4\sigma d^4\theta \ 
\left(
f{}_1{}^i{}_{a} (\bbx) \bbxi_i \bbq^a
%S_3&=& \int_{\calX} d^4\sigma d^4\theta \
+ \frac{1}{2} f_3{}^a{}_{bc}(\bbx) \bbp_a \bbq^b \bbq^c
\right),
%\nonumber \\
%\quad 
\nonumber \\
S_2
&=& \int_{\calX} d^4\sigma d^4\theta \
\frac{1}{2} f_2{}^{ab}(\bbx) \bbp_a \bbp_b,
%\nonumber \\
%S_4&=& 
\quad S_4 = \int_{\calX} d^4\sigma d^4\theta \
\frac{1}{4!} f_4{}_{abcd}(\bbx) \bbq^a \bbq^b \bbq^c \bbq^d.
%\label{4DBVaction}
\end{eqnarray*}
and denote
$\hQ_i = \sbv{S_i}{-}$ for $i=0,1,2,3,4$.
$\hQ_{0}$ has tridegree $(0,0,1)$,
$\hQ_{13}$ has $(0,1,0)$,
$\hQ_{2}$ has $(2,-1,0)$ and
$\hQ_{4}$ has $(-2,3,0)$.
%\subsection{Examples}
%\noindent
%A basis of $\calV$ is the following forms:
%\begin{eqnarray}
%%G^{(1)i} = 
%\hQ \bbx^i &=& 
%\nabla \bbx^i = 
%\bbd \bbx^i + f{}_{1}{}^i{}_{a} (\bbx) \bbq^a, 
%\nonumber \\
%%G^{(2)a} = 
%\hQ \bbq^a &=& \bbd \bbq^a
%- f_2{}^{ab}(\bbx) \bbp_b
%- \frac{1}{2} f_3{}^a{}_{bc}(\bbx) 
%\bbq{}^b \bbq{}^c
%\equiv \bbF_{q}{}^a,
%\nonumber \\
%G^{(3)}{}_a = 
%\hQ \bbp_a &=& \bbd \bbp_a
%+ f{}_1{}^i{}_{a} (\bbx) \bbxi_i 
%- f_3{}^b{}_{ac}(\bbx) \bbp_b \bbq^c
%\nonumber \\ 
%&& 
%- \frac{1}{3!} f_4{}_{abcd}(\bbx) \bbq^b \bbq^c \bbq^d
%\equiv \bbF_{p}{}_{a},
%\nonumber \\
%\hQ \bbxi_i &=& \bbd \bbxi_i 
%+ \frac{\partial f{}_1{}^j{}_{a}}{\partial \bbx^i}(\bbx) \bbxi_j \bbq^a 
%+ \frac{1}{2} \frac{\partial f_2{}^{ab}}{\partial \bbx^i}(\bbx) \bbp_a \bbp_b
%+ \frac{1}{2} \frac{\partial f_3{}^a{}_{bc}}{\partial \bbx^i}(\bbx) 
%\bbp_a \bbq^b \bbq^c
%+\frac{1}{4!} \frac{\partial f_4{}_{abcd}}{\partial \bbx^i}(\bbx) 
%\bbq^a \bbq^b \bbq^c \bbq^d
%\equiv \bbF_{\xi}{}_i,
%\nonumber
%\end{eqnarray}

\subsection{Topological Yang-Mills Theory}
%\medskip\\
%\indent
%%%%%%%%%%%%%%%%%%%%%%%%%%%%%%%%%%%%%%%%%%%%%%%%%%%%%%%%%%%%%%%%%%%%%%
%\begin{example}[Topological Yang-Mills Theory.]\label{topoYM}
%%%%%%%%%%%%%%%%%%%%%%%%%%%%%%%%%%%%%%%%%%%%%%%%%%%%%%%%%%%%%%%%%%%%%%
%\begin{example}[Topological Yang-Mills Theory.]\label{topoYM}
\noindent
{\rm 
%We take $n=3$. 
Let us consider a Lie algebra $\mathfrak{g}$
and 
the vector bundle which is
a vector space on a point.
The P-manifold over $\mathfrak{g}\to\{pt\}$
is isomorphic to $\calM = \mathfrak{g}^*[2]\oplus\mathfrak{g}[1]$
and the structure sheaf is
the polynomial algebra over $\mathfrak{g}[2]\oplus\mathfrak{g}^{*}[1]$,
$$
C^{\infty}(\calM)=S(\mathfrak{g})\otimes \bigwedge^{\cdot}\mathfrak{g}^{*}.
$$
We assume that $\mathfrak{g}$ is semi-simple.
Then the dual space $\frak{g}^{*}$ has a metric,
$(\cdot,\cdot)_{K^{-1}}$, which is
the inverse of the Killing form on $\frak{g}$.
%Then $X$ is a four dimensional manifold.
%The metric inherits the following invariant condition from
%the Killing form:
%\begin{equation}\label{la3}
%(\mathcal{L}_{p}q_{1},q_{2})_{K^{-1}}
%+(q_{1},\mathcal{L}_{p}q_{2})_{K^{-1}}=0,
%\end{equation}
%where $\mathcal{L}_{p}(-)$ is the canonical
%coadjoint action of $\mathfrak{g}$ to $\mathfrak{g}^{*}$.
%We notice that 
A Q-structure is constructed as follows:
\begin{eqnarray}
\Theta:= \theta_2 + \theta_3 =
\frac{1}{2} k^{ab}\bp_a \bp_b
+\frac{1}{2} f{}^a{}_{bc} \bp_a \bq^b \bq^c,
\label{killingQ}
\end{eqnarray}
%and $\calM$ is a QP manifold of degree $3$.
where $p_{\cdot}\in\mathfrak{g}$, $q_{\cdot}\in\mathfrak{g}^{*}$
and $k^{ab}\bp_a \bp_b:=(\cdot,\cdot)_{K^{-1}}$, and 
$f{}^{a}{}_{bc}$ is the structure constant of the Lie algebra
$\mathfrak{g}$.

(\ref{killingQ}) is obtained by
setting 
$x^i = \xi_i =0$ and
\begin{eqnarray}
f{}_1{}^i{}_{a} (x) &=& 0,
%\nonumber \\
\quad
f_2{}^{ab}(x) =
%&=& 
k^{ab}
= \delta^{ab},
%\nonumber \\
\quad
f_3{}^a{}_{bc} (x) 
%&=& 
= f{}^a{}_{bc},
%\nonumber \\
\quad
f_4{}_{abcd}(x) 
%&=& 
= 0.
%+ B
\label{4dobservable}
%\nonumber
\end{eqnarray}
in (\ref{deftheta}).
Here the orthogonal basis such that $k^{ab}= \delta^{ab}$
is taken.
%constructed from (\ref{killingQ})
A topological field theory in $4$ dimensions
is constructed 
from Equation (\ref{killingQ}):
\begin{eqnarray}
S&=&S_0 + S_2 + S_3
\nonumber \\
&=& \int_{\calX} 
d^{4}\sigma d^{4}\theta
\left(- \bbp_a \bbd \bbq^a 
+ \frac{1}{2} \bbp^a \bbp_a
+ \frac{1}{2} f{}^a{}_{bc} \bbp_a \bbq^b \bbq^c
\right).
%\nonumber \\
%&=& \int_{\calX} 
%d^{4}\sigma d^{4}\theta
%\ (- \bbp_a \bbF^a
%+ \frac{1}{2} \bbp^a \bbp_a),
\label{topoYM}
\end{eqnarray}
%where $\bbF^a = \bbd \bbq^a - \frac{1}{2} f{}^a{}_{bc} \bbq^b \bbq^c$.
%
%This is equivalent to a topological Yang-Mills theory,
%\begin{eqnarray*}
%&& S= - \frac{1}{4} \int_{\calX} 
%d^{4}\sigma d^{4}\theta
%\mu 
%\ k_{ab} \bbF^a \bbF^b,
%\end{eqnarray*}
%if we delete $\bbp_a$ by using the equations of motion.
}
%\end{example}
%%%%%%%%%%%%%%%%%%%%%%%%%%%%%%%%%%%%%%%%%%%%%%%%%%%%%%%%%%%%%%%%%
Since 
\begin{eqnarray}
&& \hQ_0^2 = \hQ_2^2 = \hQ_3^2 = 0,
\nonumber \\
&& \sbv{\hQ_k}{\hQ_l} = 0,
\qquad\qquad  \mbox{if} \qquad k \neq l,
\nonumber
\end{eqnarray}
is satisfied from Equation (\ref{tc1})-(\ref{tc3}),
$\hQ_0$, $\hQ_2$ and $\hQ_3$ define a triple complex structure
assigned tridegree $C^{(IJK)}(\hQ_{k})$ where $k, l = 0,2,3$.

The differentials with respect to $\bbq^a$ and $\bbp_a$ are 
\begin{eqnarray}
\hQ \bbq^a &=& \bbd \bbq^a
- \bbp^a
- \frac{1}{2} f{}^a{}_{bc}
\bbq{}^b \bbq{}^c,
%= \bbF^a,
\nonumber \\
\hQ \bbp_a &=& \bbd \bbp_a
- f{}^b{}_{ac} \bbp_b \bbq^c.
\nonumber
\end{eqnarray}
We expand $\bbq^a$ and $\bbp_a$ by $\theta^{\mu}$
and denote components of superfields as
\begin{eqnarray}
\bbq^a(\sigma, \theta) 
&=& \sum_{k=0}^{4} \bbq^{a(k)}(\sigma, \theta)
= \sum_{k=0}^{4} \frac{1}{k!}
\theta^{\mu(1)} \cdots \theta^{\mu(k)}
\bbq^{a(k)}_{\mu(1)\cdots \mu(k)}(\sigma)
\nonumber \\
&=& 
%\bbq^{(0)a} + \bbq^{(1)a} + \bbq^{(2)a} + \bbq^{(3)a} + \bbq^{(4)a} = 
c^a - A^a + H^{+a} + t^{+a} + v^{+a},
\nonumber \\
\bbp_a(\sigma, \theta) 
&=& \sum_{k=0}^{4} \bbp_a^{(k)}(\sigma, \theta)
= \sum_{k=0}^{4} \frac{1}{k!}
\theta^{\mu(1)} \cdots \theta^{\mu(k)}
\bbp_a^{(k)}{}_{\mu(1)\cdots \mu(k)}(\sigma)
\nonumber \\
&=& 
%\bbp^{(0)}_a + \bbp^{(1)}_a 
%+ \bbp^{(2)}_a + \bbp^{(3)}_a + \bbp^{(4)}_a = 
v_a + t_a + H_a + A^+_a + c^+_a,
\nonumber
\end{eqnarray}
The $\hQ$ differentials on the component fields are
\begin{eqnarray}
\hQ c^a 
&=& 
- v^a
- \frac{1}{2} f{}^a{}_{bc}
c^b c^c,
\nonumber \\
\hQ A^a 
&=& 
t^a
- \bbd c^a
- f{}^a{}_{bc}
A^b c^c,
\nonumber \\
\hQ v_a &=& 
- f{}^b{}_{ac} v_b c^c,
\nonumber \\
\hQ t_a &=& \bbd v_a
- f{}^b{}_{ac} (t_b c^c - v_b A^c),
\nonumber
\end{eqnarray}
which recover the BRST transformations of a topological Yang-Mills theory
\cite{Kanno:1988wm}.
\footnote{Notation in the paper \cite{Kanno:1988wm}
is $v_a = - i \phi_a$ 
and $t_a = i \psi_a$.}
%The basis of $\hQ$-exact observables 
%$\calW = \hQ \calU$ has two terms:
%are nontrivial in Equation (\ref{4dobservable}):

Let us consider $\calU = - \bbp_a \bbq^a
\in \calC^{(2,1,0)}(\hQ^{(k)})$
and a $\hQ_2$-exact cocycle:
\begin{eqnarray}
\calW_2 &=& - \frac{1}{2} \hQ_2 (- \bbp_a \bbq^a)
%\tr \ (\hQ_2 \bbq)^2
= - \frac{1}{2} (\bbp, \bbp)_{K^{-1}}
\nonumber \\
&=& - \frac{1}{2} \tr \ \left[\left(
v + t + \bbp^{(2)} + \bbp^{(3)}+ \bbp^{(4)}
\right)^2\right]
\in \calC^{(4,0,0)}(\hQ^{(k)}),
\nonumber
\end{eqnarray}
is a $\hQ_{\inte}$-cohomology class
because $\hQ_3 \calW =0$ from 
$\rho=0$ in the identity $(A4)$,
where $\tr$ is a trace with respect to the Killing form.
%,
If all the antifields $\bbp^{(3)}$ and $\bbp^{(4)}$ are set to zero
and the equation of motion for $\bbp^{(2)}_a$ is used:
\begin{eqnarray}
\bbp^{(2)}_a &=& - F^a =
- \bbd A^a
- \frac{1}{2} f_3{}^a{}_{bc}
%(\bbx) 
A{}^b A{}^c,
\nonumber 
\end{eqnarray}
$\calW_2$ becomes the following form:
\begin{eqnarray}
\calW_2 
&=& - \frac{1}{2} \tr \left[\left(v
+ t - F
\right)^2 \right]
= - \frac{1}{2} \tr \left[\left(- i \phi
+ i \psi - F
\right)^2 \right]
\nonumber \\
&=& 
- \frac{1}{2} \tr F^2 
+ i \tr \psi F
+ \tr \left(\frac{1}{2} \psi^2 - i \phi F \right)
- \tr \psi \phi 
+ \frac{1}{2} \tr \phi^2,
\nonumber
\end{eqnarray}
which is an original Donaldson polynomial invariant
in the topological Yang-Mills theory 
\cite{Donaldson}\cite{Witten:1988ze}\cite{Kanno:1988wm}.
Here $v_a = - i \phi_a$ and $t_a = i \psi_a$.
%$(\hQ_0 + \hQ_1^{(3)}) \calW_2 =0$ is satisfied.
%\medskip\\
%\indent
More generally, 
the trace of $N$ times of $\hQ_2$ exact cocycles, 
$- \bbp^a = \hQ_2 \bbq^a$:
\begin{eqnarray}
\calW_N = c_N \tr \ (- \bbp^a)^N
= c_N \tr \ (\hQ_2 \bbq^a)^N
\in \calC^{(2N,0,0)}(\hQ^{(k)}),
\nonumber
\end{eqnarray}
is a $\hQ_{\inte}$-invariant polynomial
$\hQ_{\inte} \calW_N =0$,
because 
$\hQ_3 \hQ_2 \bbq^a = -[\bbp, \bbq]^a$.
%= -f^{b}{}_{ac} \bbp_b \bbq^c$.
Here $c_N$ is a normalization constant.
%$\calW_N$ is a $\hQ$ coclyce.
$\calW_N$ is expanded by the order of $\theta$'s 
as
\begin{eqnarray}
\calW_N = \sum_{k=0}^{4} \calW^{(N)}_k (\sigma, \theta)
= \sum_{k=0}^{4} 
\theta^{\mu(1)}\cdots \theta^{\mu(k)}
\calW^{(N)}_{k, \mu(1) \cdots \mu(k)}(\sigma).
\nonumber
\end{eqnarray}
By substituting this expansion to the equation 
$\hQ_{\inte} \calW_N =
(\hQ - \hQ_0) \calW_N=0$
%(\ref{QWN}), 
and considering descent equations,
the following theorem is obtained
from the theorem (\ref{kcocycle}):
\begin{thm}
The integration of $\calW^{(N)}_k$
on a $k$-cycle $\gamma_k$ of $\calX$,
\begin{eqnarray}
\calO^{(N)}_{k} = \int_{\gamma_k} \mu_k \ \calW_{N}
= \int_{\gamma_k} \mu_k \ \calW^{(N)}_k,
\nonumber
\end{eqnarray}
is a $\hQ$-cohomology class.
\end{thm}
%by applying the theorem \ref{q1cohomology} to $\calW^{(N)}_k$.

%%%%%%%%%%%%%%%%%%%%%%%%%%%%%%%%%%%%%%%%%%%%%%%%%%%%%%%%%%%%%%%%%%%%%%%%%%%%%
Since $\calW_N$ is $\hQ_2$-exact,
there exists 
%an analogue of 
a Chern-Simons form $\calI_N$ 
such that 
\begin{eqnarray}
\calW_N = \hQ_2 \calI_N.
\nonumber
\end{eqnarray}
%Let us consider Example \ref{topoYM},
%the topolgical Yang-Mills theory, 
%In the topolgical Yang-Mills theory,
%%%%%%%%%%%%%%%%%%%%%%%%%%%%%%%%%%%%%%%%%%%%%%%%%%%%%%%%%%%%%%%%%%%%%%%%%%%%%
%This cocylce is not $\hQ_1$-exact.
%Let us take
%$\calU = \bbe_{a(i)}$.
%Then 
%$$
%\calW 
%%= \hQ \calG 
%= \hQ \bbe_{a(i)} 
%= \bbd \bbe_{a(i)} 
%+ \sbv{S_1}{\bbe_{a(i)}}
%= \bbd \bbe_{a(i)} 
%- \frac{S_1 \rd}{\partial \bbe_{b(j)}} \bomega_{b(j)a(i)}
%$$
%Especially, 
%$\calG = \bbx^{a(0)} = \bbx^{i}$,
%where we denote $a(0)=i$.
%Then
%$$
%\calF = \hQ \bbx^{i} = \bbd \bbx^i 
%- \frac{S_1 \rd}{\partial \bbxi_j}\frac{\ld}{\partial \bbx^j} \bbx^i
%= \bbd \bbx^i 
%- \frac{S_1 \rd}{\partial \bbxi_i}.
%$$
%\medskip\\
%\indent

An original action of the topological Yang-Mills theory
is derived by carrying out a gauge fixing 
of the AKSZ action (\ref{topoYM}) 
and restricting the theory to a Lagrangian submanifold.
In order to fix gauge, two antighosts and 
two Nakanishi-Lautrap Lagrange multiplier fields
are introduced.
$\lambda^a$ is a Grassmann even scalar field of ghost number $-2$
and $\eta^a$ is a Grassmann odd scalar field of ghost number $-1$.
$\lambda^{+}_a$ and $\eta^{+}_a$ are their Hodge dual antifields.
$\chi^a$ is a Grassmann odd two form (a function of $\theta^2$) 
of ghost number $-1$
and $\zeta^a$ is a Grassmann even two form of ghost number $0$.
$\zeta^{+}_a$ and $\chi^{+}_a$
are their Hodge dual antifields.
A gauge fixing term is added 
to the action $S$:
\begin{eqnarray}
S_{GF} = \int_{\calX} d^4\sigma d^4\theta
(\eta^a \lambda^{+}_a 
%+ f^a{}_{bc} \eta^{+b} v_a \lambda^c
+ \zeta^a \chi^{+}_a
).
\end{eqnarray}
and a following gauge fixing fermion term is considered:
\begin{eqnarray}
\Psi = \int_{\calX} d^4\sigma d^4\theta
\left(\lambda^a (D_A * t)_a + f^a{}_{bc} \eta^b 
v_a \lambda^c
+ \chi^a \left(H_a - * H_a
+ \frac{1}{2} f_{abc} c^b (\chi^c - * \chi^c)
\right)
\right),
\end{eqnarray}
where $(D_A \Phi)_a = \bbd \Phi_a
+ f^b{}_{ac} A^c \Phi_b$ and
$(D_A \Phi)^a = \bbd \Phi^a
+ f^a{}_{bc} A^b \Phi^c$
are covariant derivatives.
The gauge fixing conditions are
\begin{eqnarray}
\lambda^+_a &=& \frac{\partial \Psi}{\partial \lambda^a}
= (D_A * t)_a,
\nonumber \\
\chi^+_a &=& \frac{\partial \Psi}{\partial \chi^a}
= \tilde{H}_a = H_a - * H_a,
\nonumber \\
A^+_a &=& \frac{\partial \Psi}{\partial A^a}
= f^c{}_{ba} \lambda^b  * t_c,
\nonumber \\
H^{+a} &=& \frac{\partial \Psi}{\partial H_a}
= \tilde{\chi}^a = \chi^a - * \chi^a,
\nonumber \\
t^{+a} &=& \frac{\partial \Psi}{\partial t_a}
= * (D_A \lambda)^a,
\nonumber \\
%\eta^{+}_a &=& \frac{\partial \Psi}{\partial \eta^a}
%= f^b{}_{ac} v_b \lambda^c,
%\nonumber \\
%v^{+a} &=& \frac{\partial \Psi}{\partial v_a}
%= f^a{}_{bc} \eta^b \lambda^c,
%\nonumber \\
c^{+}_a &=& - \frac{1}{2} f^a{}_{bc} \tilde{\chi}^b 
\tilde{\chi}^c,
\nonumber \\
c^a &=& 0,
\nonumber 
\end{eqnarray}
where 
$\tilde{H}_a = H_a - * H_a$ and 
$\tilde{\chi}^a = \chi^a - * \chi^a$
are selfdual parts of each field.
The gauge fixed action becomes 
\begin{eqnarray}
S + S_{GF}| 
&=&
\int 
\biggl(\frac{1}{2} H^2 + H_a F^a + \zeta^a \tilde{H}_a
- v_a D_A * D_A \lambda^a + \eta^a D_A * t_a
- t_a D_A \tilde{\chi}^a
\nonumber \\
&& \qquad 
%+ \frac{1}{2} f^a{}_{bc} v_a \tilde{\chi}^b \tilde{\chi}^c
-  f^c{}_{ab} \lambda^a t_b * t_c
%+  f^a{}_{bc} v_a \eta^b \eta^c 
%+  f^a{}_{bc} f^{db}{}_{e} v_d \lambda^e v_a \lambda^c
\biggr).
\nonumber
\end{eqnarray}
If we delete $H_a$ by the equations of motion, we obtain 
the action of a topological Yang-Mills theory 
\footnote{The original notation in the
paper \cite{Witten:1990bs} is $v_a = \phi_a$ 
and $t_a = - \psi_a$.}
\cite{Witten:1990bs}:
\begin{eqnarray}
S + S_{GF}| 
&=&
\int 
\biggl(
- \frac{1}{2} \tilde{F}^a \tilde{F}_a 
- v_a D_A * D_A \lambda^a + \eta^a D_A * t_a
- t_a D_A \tilde{\chi}^a
%\nonumber \\
%&& \qquad 
%+ \frac{1}{2} f^a{}_{bc} v_a \tilde{\chi}^b \tilde{\chi}^c
-  f^c{}_{ab} \lambda^a t_b * t_c
%+  f^a{}_{bc} v_a \eta^b \eta^c 
%+  f^a{}_{bc} f^{db}{}_{e} v_d \lambda^e v_a \lambda^c
\biggr).
\nonumber
\end{eqnarray}

%%%%%%%%%%%%%%%%%%%%%%%%%%%%%%%%%%%%%%%%%%%%%%%%%%%%%%%%%%%%%%%%%%%%%
\section{Other Examples}
%%%%%%%%%%%%%%%%%%%%%%%%%%%%%%%%%%%%%%%%%%%%%%%%%%%%%%%%%%%%%%%%%%%%%%
%\begin{example}[Topological Yang-Mills Theory.]\label{topoYM}
%%%%%%%%%%%%%%%%%%%%%%%%%%%%%%%%%%%%%%%%%%%%%%%%%%%%%%%%%%%%%%%%%%%%%
%%%%%%%%%%%%%%%%%%%%%%%%%%%%%%%  SEC. 4 %%%%%%%%%%%%%%%%%%%%%%%%%%%%%%
%%%%%%%%%%%%%%%%%%%%%%%%%%%%%%%%%%%%%%%%%%%%%%%%%%%%%%%%%%%%%%%%%%%%%
\subsection{$\theta_{4}=0$}
\noindent
In this case, the bracket (\ref{braee}) satisfies
(A0), (A1) and the Jacobi identity.
Therefore, the bundle $E\to M$ becomes a Lie algebroid:
\begin{definition}
(\cite{Mackenzie})
A Lie algebroid over a manifold $M$ is a vector bundle
$E \rightarrow M$ with a Lie algebra structure on the 
space of the sections $\Gamma E$ defined by the 
bracket $[e_1, e_2]$ for $e_1, e_2 \in \Gamma E$
and an anchor map
$\rho: E \rightarrow TM$ satisfying (A0) and (A1).
\end{definition}
%the following properties:
%\begin{eqnarray}
%\label{liealgdef1}\rho[e_1, e_2]&=&[\rho(e_1), \rho(e_2)]
%\label{liealgdef2}[e_1, f e_2]&=&f [e_1, e_2] + (\rho(e_1) f) e_2,
%\end{eqnarray}
%where $e_1$ and $e_2$ are sections of $E$ and 
%$f \in C^{\infty}(M)$.\\
\indent
Let us take $\{ e_a \}$ as a local basis of $\Gamma E$ and
a local expression of an anchor map
 $\rho(e_a) = f^i{}_{1a}(x) \frac{\partial}{\partial x^i}$
and a Lie bracket
$[e_b, e_c] = f_3{}^a{}_{bc}(x) e_a$.
The Q-structure $\Theta$ associated with
the Lie algebroid $E$
is defined as a function on $T^{*}[3]E[1]$,
\begin{eqnarray}
\Theta:=\theta_{123}
:=f{}_1{}^i{}_{a} (\bx) \bxi_i \bq^a
+\frac{1}{2} f_2{}^{ab}(\bx) \bp_a \bp_b
+\frac{1}{2} f_3{}^a{}_{bc}(\bx) \bp_a \bq^b \bq^c,
\label{liealgebroid}
\end{eqnarray}
which is globally well-defined.
Here $\frac{1}{2} f_2{}^{ab}(\bx) \bp_a \bp_b$ 
is a symmetric pairing on $E^*$.
%Conversely, if we consider $\Theta:=\theta_{13}$,
%the classical master equation
%induces the Lie algebroid structure on $E$.
%We note that the equations 
%(\ref{liealgdef1}) and (\ref{liealgdef2}) are the 
%same as 
%the conditions (A0) and (A1).

The AKSZ construction derives the Q-structure
\begin{eqnarray}
S = S_0 + S_{13} +S_2.
\end{eqnarray}
%from (\ref{liealgebroid}).
$S$ defines a Chevelley-Eilenberg differential 
%of a Lie algebroid
on $\Map(\calX, \calM)$.
If we construct $\calW$ such that $Q_{\inte} \calW =0$,
$\mu_{*k} \calW$ is a Chevelley-Eilenberg cohomology class
on a Lie algebroid $E$.

The theory has a triple complex structure because
\begin{eqnarray}
&& \hQ_0^2 = \hQ_{13}^2 = \hQ_2^2 = 0,
\nonumber \\
&& \sbv{\hQ_k}{\hQ_l} = 0,
\qquad\qquad  \mbox{if} \qquad k \neq l,
\nonumber
\end{eqnarray}
where $k,l= 0, 13, 2$.
If we consider $\calW_2 = (\hQ_2 \bbq, \hQ_2 \bbq)$,
$\hQ_{\inte} \calW_2 = 0$ since 
$\hQ_{2} \calW_2 = \hQ_{13} \calW_2 = 0$.
Therefore $\mu_{*} \calW_2 $ is a $\hQ$-cohomology class
and $\calW_2$ is a generalization of a Donaldson polynomial
for a Lie algebroid.

%%%%%%%%%%%%%%%%%%%%%%%%%%%%%%%%%%%%
\subsection{$\theta_{2}=0$}
%%%%%%%%%%%%%%%%%%%%%%%%%%%%%%%%
In this case, A Q-structure $\hQ$ is decomposed 
to $\hQ = \hQ_0 + \hQ_{13} + \hQ_4$
and
\begin{eqnarray}
&& \hQ_0^2 = \hQ_{13}^2 = \hQ_4^2 = 0,
\nonumber \\
&& \sbv{\hQ_k}{\hQ_l} = 0,
\qquad\qquad  \mbox{if} \qquad k \neq l.
\nonumber
\end{eqnarray}
Therefore $\hQ_{13}$ and $\hQ_4$ define
a triple complex structure.
%from (\ref{tc1})-(\ref{tc3}).

An important cocycle is 
$\calW = \hQ_4 (\frac{1}{4} \bbp_a \bbq^a)$,
This is a $\hQ_{\inte}$ invariant class and its integration defines a
$\hQ$-cohomology class.
Since $\calW = \frac{1}{4!} f_4{}_{abcd}(\bbx) \bbq^a \bbq^b \bbq^c \bbq^d$,
$\int_{\calX} \mu \ \calW \in H^{4}_{dR}(\bigwedge^{\cdot}E^{*},\delta)$,
which classifies a QP-structure of degree $3$ in $\theta_2 =0$ case.

%%%%%%%%%%%%%%%%%%%%%%%%%%%%%%%%%%%%%%%%%%%%%%%%%%%%%%%%%%%%%%%%%%%%%
%%%%%%%%%%%%%%%%%%%%%%%%%%%%%%   SEC  6    %%%%%%%%%%%%%%%%%%%%%%%%%%
%%%%%%%%%%%%%%%%%%%%%%%%%%%%%%%%%%%%%%%%%%%%%%%%%%%%%%%%%%%%%%%%%%%%%
%\section{Conclusions and Discussion}
%\noindent

%%%%%%%%%%%%%%%%%%%%%%%%%%%%%%%%%%%%%%%%%%%%%%%%%%%%%%%%%%%%%%%%%%%%%
%%%%%%%%%%%%%%%%%%%%%%%%%%%%%%   APPENDIX    %%%%%%%%%%%%%%%%%%%%%%%%
%%%%%%%%%%%%%%%%%%%%%%%%%%%%%%%%%%%%%%%%%%%%%%%%%%%%%%%%%%%%%%%%%%%%%
%\section*{Appendix}
%\noindent
%

%%%%%%%%%%%%%%%%%%%%%%%%%%%%%%%%%%%%%%%%%%%%%%%%%%%%%%%%%%%%%%%%%%%%%
%%%%%%%%%%%%%%%%%%%%%%%%%%%%   ACKNOWLEGE    %%%%%%%%%%%%%%%%%%%%%%%%
%%%%%%%%%%%%%%%%%%%%%%%%%%%%%%%%%%%%%%%%%%%%%%%%%%%%%%%%%%%%%%%%%%%%%
\section*{Acknowledgments}

The author would like to thank K.Koizumi and K.Uchino 
for valuable discussions and comments.
This work is supported by Maskawa Institute, Kyoto Sangyo University.

\newcommand{\bibit}{\sl}
%%%%%%%%%%%%%%%%%%%%%%%%%%%%%%%%%%%%%%%%%%%%%%%%%%%%%%%%%%%%%%%%%%%%%
%%%%%%%%%%%%%%%%%%%%%%%%%%%%%%%  Refs. %%%%%%%%%%%%%%%%%%%%%%%%%%%%%%
%%%%%%%%%%%%%%%%%%%%%%%%%%%%%%%%%%%%%%%%%%%%%%%%%%%%%%%%%%%%%%%%%%%%%
%\newpage
%NEW MACRO FOR BIBLIOGRAPHY

%\section*{References}
%\noindent

\vfill\eject
\end{document}